\documentclass[usegraphicx,usenatbib,useAMS]{mn2e}

\usepackage{multirow}
\usepackage{xcolor}
\usepackage{aas_macros}
\usepackage{amssymb}
\usepackage{xspace}
\usepackage[normalem]{ulem}
\usepackage{enumitem}
\usepackage{amsmath}
\usepackage{algorithmicx}
\usepackage{algpseudocode}

\newcommand{\SUBFIND}{\textsc{subfind}\xspace}

\title[Halo spin evolution model]
{A random walk model for the evolution of the halo spin vector}

\author[Hou et al.]  {Jun Hou\thanks{\scriptsize E-mail:
    junhou@shnu.edu.cn (JH); zjluo@shnu.edu.cn (ZJL); cedric.lacey@durham.ac.uk (CGL)},$^{1}$ Zhijian Luo\footnotemark[1],$^1$ Cedric G.\ Lacey\footnotemark[1]$^2$
  \\
  $^{1}$Shanghai Key Lab for Astrophysics, Shanghai Normal University, 
  100 Guilin Road, Shanghai 200234, China
  \\  
  $^{2}$Institute for Computational Cosmology, Department of Physics, University of Durham, South Road, Durham DH1 3LE, UK}

\begin{document}
	
	\maketitle
	
	\begin{abstract}  
		We follow the spin vector evolutions of well resolved dark matter haloes (containing more than $300$ particles) in merger tree main branches from the Millennium and Millennium-II N-body simulations, from $z\sim 3.3$ to $z=0$. We find that there seems to be a characteristic plane for the spin vector evolution along each main branch. In the direction perpendicular to it, spin vectors oscillate around the plane, while within the plane, spin vectors show a coherent direction change as well as a diffusion in direction (possibly corresponds to a Gaussian white noise). This plane may reflect the geometry of surrounding large-scale structures. We also construct a simple stochastic model in which halo spin vector evolution is assumed to be driven by accretion of halo mass and angular momentum. This model can reproduce major features of the results from N-body simulations. 
	\end{abstract}
	
	
	\section{Introduction}
	\label{sec:introduction}
	
	The spin vector of a dark matter halo is a fundamental property, and therefore its various features and evolution are a window for understanding non-linear structure formation. It is well accepted that the magnitude of halo spin, when expressed in terms of the dimensionless spin parameter $\lambda$, follows a log-normal distribution, which depends at most weakly on halo mass, and is not very sensitive to redshift \citep[e.g.][]{lambda,lambda_distri_Vitvistska,lambda_distri_Bailin,Bett_2007}. \citet{lambda_evolution_Kim} further shows that the evolution of the magnitude of the spin is approximately a Markov random walk, and is mainly correlated to the halo mass accretion history, although there are secondary correlations with redshift and environment.
	
	For the spin directions, there have been many investigations of their correlations with halo shapes and large-scale structures, and how these correlations change with redshifts (see e.g.\ \citet{spin_filament_Ganeshaiah3} and references therein). Works of this kind have revealed many detailed and complex features of these correlations and their evolutions. On the other hand, the evolution of spin directions, namely how the spin direction of a halo changes relative to that of its progenitor halo, has received less attention. \citet{spin_change_Padilla} investigated the angle between halo spins before and after halo mass changes, and derived the probability distributions of this angle in different redshift ranges. \citeauthor{spin_change_Padilla} found that on average, this angle tends to be large at higher redshifts. This work was then expanded in \citet{halo_spin_Contreras}, which considered how halo spin directions change with time from $z\sim 1$ to $z=0$. \citeauthor{halo_spin_Contreras} not only considered the angle between halo spins at different times, but also the relative orientation of the planes containing these angles. Together, these two quantities give a description of halo spin direction change in 3D space, although it is somewhat indirect. \citeauthor{halo_spin_Contreras} also investigated the correlations between spin direction changes and other quantities such as halo mass changes, redshifts, spin magnitudes etc. But one should note that \citet{halo_spin_Contreras} mainly focuses on the median evolution of spin directions.
	
	This work tries to extend the work of \citet{halo_spin_Contreras}. We adopt a more straightforward description of spin directions, i.e.\ the azimuthal and polar angles of a specially chosen spherical coordinate system. We then consider the spin direction changes beyond the median evolution through investigating the autocorrelations of these two angles, and the cross-correlation between each pair of them and spin magnitude. We also extend the redshift range, and in this work we consider spin direction evolution between $z\sim 3.3$ and $z=0$.
	
	The estimation of galaxy angular momenta in many simplified galaxy formation models is based on halo spins \citep[e.g.][]{cole2000,lgalaxies_2015,galform_2016,shark}. These models adopt either N-body or Monte Carlo merger trees to follow dark matter halo growth. For the former, reliable merger trees can include small haloes that are resolved with only a few tens of particles, while reliable halo spin vectors can only be measured for haloes with at least a few hundreds of particles. A better understanding of spin vector evolution would help to fill this gap. On the other hand, Monte Carlo halo merger trees do not contain spin information, and some additional spin vector assignment algorithm is required. A better understanding of spin vector evolution would help in the construction of such algorithms. All these would eventually improve the quality of the model predictions of galaxy properties.  
	In this work we construct a stochastic, random walk model for the magnitude and direction of the spin vector, which can reproduce the major features of the spin vector evolution we obtained from N-body simulations. This model can be used in future simplified galaxy formation models.
	
	Previously \citet{Benson2020} also constructed a random walk model for spin vector evolution, based on several detailed assumptions about the orbital properties of haloes being accreted. Our model does not involve assumptions of this kind, and therefore is much simpler than that in \citeauthor{Benson2020} When compared with results measured from N-body simulations, \citeauthor{Benson2020} only considered haloes with $z=0$ mass larger than $3\times 10^{13}\,{\rm M}_\odot$, because of the limited resolution of the Millennium simulation used in that work. This work extends the mass range of this comparison through adopting both Millennium and Millennium-II simulations, with the latter having resolution $125$ times higher than the former. Although the model in \citeauthor{Benson2020} is capable of predicting the full direction information of a spin vector, in that work, the comparison with N-body simulations is limited to angles between spin vectors of haloes at different times, which only contain part of the direction information, while in this work a full comparison is done through the previously mentioned specially chosen spatial frames.
	
	This paper is organized as follows. Section~\ref{sec:halo_sample} introduces the N-body simulations and halo samples used in this work. Section~\ref{sec:halo_spin} describes how halo spin vectors are measured and how the spatial frames for describing their directions are constructed. Our main results derived from N-body simulations are given in Section~\ref{sec:j_m_relation} to \ref{sec:cross_correlations}, and a simple stochastic model for spin evolution is introduced in Section~\ref{sec:simple_model_description}, followed by a simple comparison between our model and the model in \citet{Benson2020}. Finally, a summary is given in Section~\ref{sec:summary}.


\section{Simulation data}\label{sec:data}
\subsection{Simulations and halo samples}\label{sec:halo_sample}
In this work we use the Millennium \citep{millennium} and Millennium-II \citep{millennium2} pure dark matter N-body simulations. Both of these simulations are run with the following cosmological parameters: $\Omega_{\rm tot}=1, \Omega_{\rm M}=0.25, \Omega_{\rm \Lambda}=0.75, h=0.73, \sigma_{8}=0.9, n_{s}=1$, and they have the same number of particles, i.e.\ $2160^3$. The Millennium simulation is run in a cube with side length $500\,h^{-1}{\rm Mpc}$, and its mass resolution is $1.18\times 10^9\,{\rm M}_\odot$. The Millennium-II simulation is run in a cube with side length $100\,h^{-1}{\rm Mpc}$, and hence its mass resolution is $125$ times better than the Millennium simulation, and reaches $9.43\times 10^6\,{\rm M}_\odot$.

The structures formed in these simulations are first identified through the friends-of-friends (FOF) algorithm \citep{FOF}, and then each of these FOF groups is further spilt into subgroups through \SUBFIND \citep{subfind}. Halo merger trees are constructed based on these subgroups through the Dhalo algorithm \citep{Dhalo_Helly,Dhalo_Jiang}. This algorithm first generates merger trees of subgroups through matching the most bounded particles at different snapshots. Then, it groups different subgroups into Dhalos by examining their separations: if one subgroup lies within twice the half-mass radius of another subgroup, these these two are identified as being in the same Dhalo. Once a subgroup is identified as a part of a Dhalo, it is always considered to belong to this Dhalo. With Dhalos being identified, subgroup merger trees are assembled into Dhalo merger trees.

The original FOF groups suffer from several artificial effects \citep[e.g.][]{Dhalo_Jiang}. First, two distinct structures may temporarily be gathered into a single FOF group through artificial low density bridges. This effect causes an FOF group to artificially first gain and then lose mass. Second, small structures may oscillate around a major structure before finally merge into it, and this causes an FOF group to artificially first lose mass and then gain this mass back after one or few snapshots. As pointed out in \citet{halo_spin_Contreras}, these artificial mass changes accompany artificial halo spin changes, and should be corrected before reliable halo spin evolution can be derived. 

The construction of Dhalos can largely avoid the above-mentioned two effects. Dhalos are grouped through checking subgroup separations, and those connected through artificial low density bridges would not be put into a single Dhalo. Further, a Dhalo member is always considered to be in this Dhalo, no matter it temporarily leaving the associated FOF group or not. With these treatments, Dhalo masses almost always evolve monotonously. 

In this work we construct three halo samples based on Dhalo merger trees. We select $z=0$ haloes with masses in three ranges: $[10^{11}\,{\rm M}_\odot,10^{11.5}\,{\rm M}_\odot]$, $[10^{12}\,{\rm M}_\odot,10^{12.5}\,{\rm M}_\odot]$ and $[10^{13.5}\,{\rm M}_\odot,+\infty)$. The lower mass boundary, $10^{11}\,{\rm M}_\odot$, is selected to ensure these haloes to have well resolved ($\geq 300$ particles) progenitors up to $z\sim3$. These three ranges approximately represent the low, intermediate and high density environment of structure formation. We postpone a more detailed study of the environmental effects on spin evolution to future works. Haloes in the first two ranges are selected from the Millennium-II simulation, because there they are better resolved than in the Millennium simulation, while haloes in the above third range are from the Millennium simulation, because its large simulation volume allow a statistically meaningful sample of these massive rare haloes. At $z=0$, there are respectively $22727$, $3068$ and $19692$ haloes in these three samples. We then follow the halo spin evolution along the main branches of the merger trees of these haloes. Here the main branch of a merger tree is the branch formed by the most massive progenitor haloes.

\subsection{Spin direction measurement}\label{sec:measure_direction}
\subsubsection{Deriving spin vectors}\label{sec:halo_spin}
In this work we derive halo spin evolution through spin vectors of Dhalos along the main branch of a given merger tree. As further described in \S\ref{sec:halo_sample}, the construction of Dhalo can largely reduce the impact on spin evolution from artificial effects in structure identification. We also limit our spin measurements to Dhalos with at least $300$ particles, because according to \citet{Bett_2007}, reliable spins can be derived only for those well resolved haloes. Under this limitation, main branches stretch to $z\sim 3.3$.

The spin vector of a Dhalo is derived through adding angular momenta of all particles in this halo. The reference point of these angular momenta is the centre of mass of this halo, and the particle velocities used to calculate these angular momenta are those with respect to this reference point. We use the following dimensionless vector to describe halo spin:
\begin{equation}
\boldsymbol{\lambda}=\frac{\boldsymbol{j_{\rm halo}}}{\sqrt{2}r_{\rm vir}v_{\rm vir}},
\label{eq:lambda_vector}
\end{equation}
where $\boldsymbol{j_{\rm halo}}=\boldsymbol{J}_{\rm halo}/M_{\rm halo}$ is the halo specific angular momentum, $v_{\rm vir}=\sqrt{GM_{\rm halo}/r_{\rm vir}}$, and $r_{\rm vir}$ is the halo virial radius. $r_{\rm vir}$ is derived from halo mass, $M_{\rm halo}$, and halo redshift, $z_{\rm halo}$, through $r_{\rm vir}=\sqrt[3]{3M_{\rm halo}/[4\pi\Delta_{\rm vir}(z_{\rm halo})\rho_{\rm cirt}(z_{\rm halo})]}$, where $\rho_{\rm crit}(z)$ is the cosmic critical density at redshift $z$, and $\Delta_{\rm vir}(z)$ is the spherical collapse overdensity at this redshift. In this work, we adopt the simple fitting formula for $\Delta_{\rm vir}(z)$ from \citet{spherical_overdensity_Eke} and \citet{spherical_overdensity_Bryan}. The magnitude of this vector, $\lambda\equiv|\boldsymbol{\lambda}|$, is the familiar dimensionless spin parameter introduced in \citet{lambda}.

\subsubsection{Constructing the frame for spin direction description}\label{sec:frame_construction}
To straightforwardly describe the direction of $\boldsymbol{\lambda}$ in 3D space, a spatial frame is necessary. Because of the diverse orientations of structures in the simulation volume, adopting a fixed frame for all haloes would not be efficient to extract statistical features of spin direction evolutions. Therefore, we construct a different frame for each of the merger tree main branches. 

We construct this kind of frames in three steps. Let $\boldsymbol{\lambda}_{\rm i}$ be the spin vector of a halo at the $i$-th snapshot, and $\boldsymbol{\lambda}_{\rm i+1}$ the spin vector of its descendant halo at the next snapshot. Usually these two vectors are not co-linear and we denote the unit vector parallel to $\boldsymbol{\lambda}_{\rm i}\times \boldsymbol{\lambda}_{\rm i+1}$ as $\hat{\boldsymbol{n}}_{\rm i}$. Clearly, $\hat{\boldsymbol{n}}_{\rm i}$ is a normal vector of the plane spanned by these two halo spins. In general, the spin direction only has small changes between adjacent snapshots\footnote{this is not true when major mergers happen, but we checked that they are not frequent along a typical main branch, and therefore should not spoil the averaging in equation~(\ref{eq:average_norm_v}). }, therefore, $\boldsymbol{\lambda}_{\rm i}$ usually transfers to $\boldsymbol{\lambda}_{\rm i+1}$ through rotating by an angle with absolute value smaller than $180^\circ$. The definition of $\hat{\boldsymbol{n}}_{\rm i}$ ensures this rotation angle is positive. For a given merger tree main branch, starting from its tip, this kind of unit normal vector can be derived for each pair of adjacent snapshots until one snapshot in this pair reaches $z=0$. Then a main branch gives a series of unit normal vectors, and we denote their average as
\begin{equation}
\bar{\boldsymbol{n}}=\frac{1}{N}\sum_{j=i_{\rm initial}}^{i_{\rm final}} \hat{\boldsymbol{n}}_{j},
\label{eq:average_norm_v}
\end{equation}
where the summation is from the earliest snapshot ($i_{\rm initial}$-th) to the latest one ($i_{\rm final}$-th) along the main branch, and $N$ is the total number of vectors in this summation. Usually $\bar{\boldsymbol{n}}$ is not $\boldsymbol{0}$, and then it can be viewed as a normal vector of a new plane. We call this plane the average plane, because its normal vector is from the average process described in equation~(\ref{eq:average_norm_v}). We set the z-axis of our spatial frame to be parallel to $\bar{\boldsymbol{n}}$. We then set the x-axis of this frame to be parallel to the component of the initial halo spin (spin of the halo at the tip of a given main branch) that lies within the average plane. The y-axis is set such that the three axes form a right-handed Cartesian frame. 

\begin{figure}
	\centering
	\includegraphics[width=0.48\textwidth]{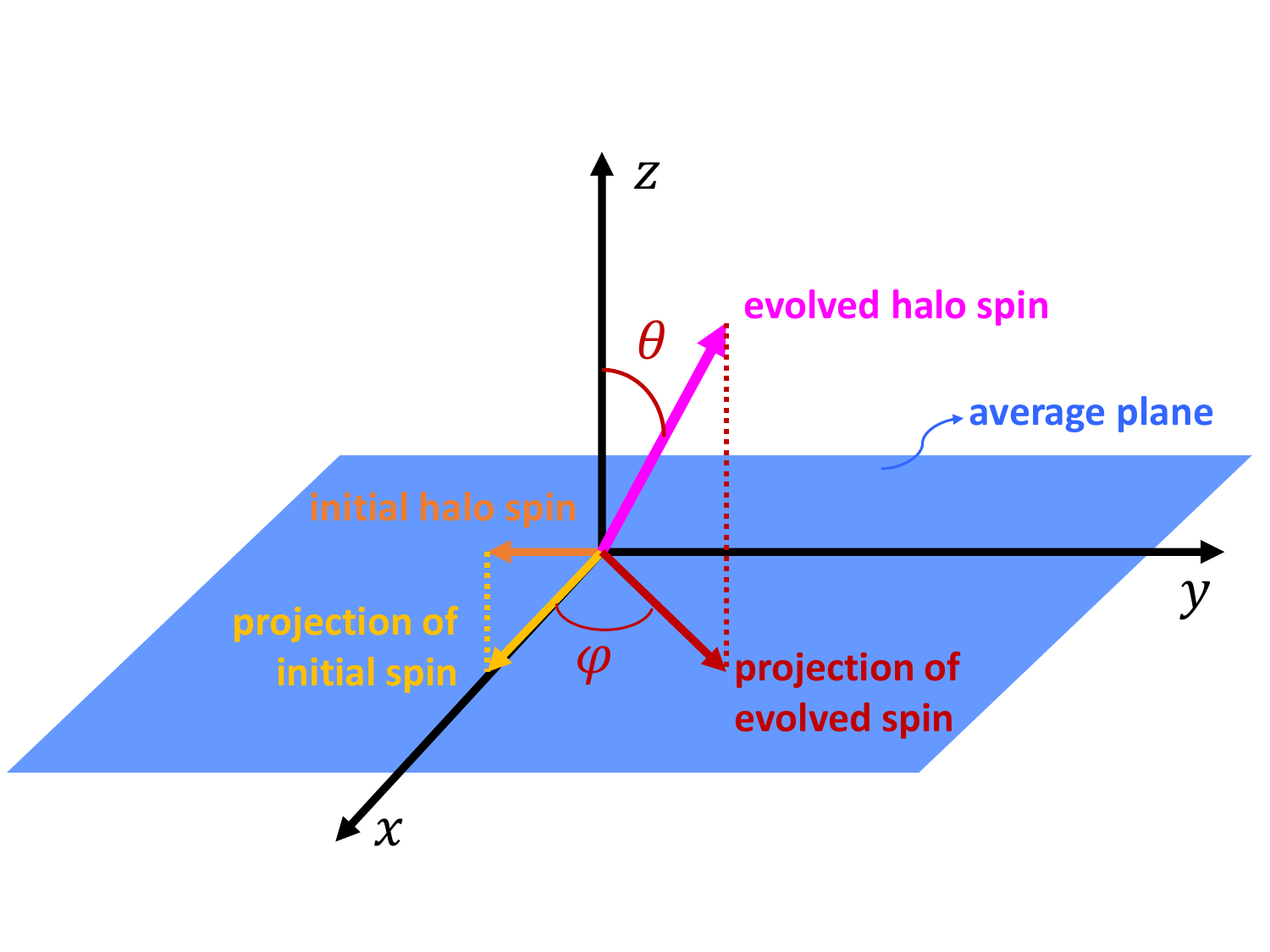}
	\caption{An illustration of the spatial frame and associated angular coordinates for describing halo spin directions. Details of the frame construction and coordinate definitions are given in \S\ref{sec:frame_construction}.}	
	\label{fig:frame}
\end{figure}

A spherical coordinate system is attached to this Cartesian frame through a standard way: the polar axis is the z-axis, and the prime meridian plane is the half plane spanned by the z-axis and the positive part of x-axis. The two angular coordinates of this spherical coordinate system, i.e. the azimuthal angle $\varphi$ and the polar angle $\theta$, provide an efficient description of directions of $\boldsymbol{\lambda}$. In this work, we choose the following value ranges for these two angle coordinates: $\theta\in[0^\circ, 180^\circ]$, $\varphi\in(-180^\circ,180^\circ]$. The Cartesian frame and associated two angular coordinates are summarized in Fig.~\ref{fig:frame}.

If there is a coherent direction change of $\boldsymbol{\lambda}$ along a main branch, then in the frame constructed above, it should manifest itself as an coherent increase of $\varphi$. This feature stems from the definition of $\hat{\boldsymbol{n}}_{\rm i}$ and that of the normal vector of the average plane.

Along each merger tree main branch, we measure the $\varphi$ and $\theta$ of halo spins, and then we combine all measurements from a given halo sample to derive statistical features of halo spin direction evolution.

\subsubsection{Approximate periodic correction for $\varphi$}\label{sec:phi_correction}
As a coordinate, $\varphi$ has finite value range, and here it is $(-180^\circ,180^\circ]$. When the direction change of $\boldsymbol{\lambda}$ leads $\varphi$ to move out of this value range from one side, $\varphi$ then re-enters it from the other side. This behaviour causes a large jump in $\varphi$, and masks its essentially continuous evolution. Therefore, this effect needs to be corrected. Because N-body simulations only provide $\varphi$ at a series of discrete output times, instead of $\varphi$ as a function of time, this correction can only be approximate.

More specifically, we denote a $\varphi$ evolution track extracted from N-body simulations as $\{ \varphi_i \}_{\rm i=1}^{\rm N}$, where $\varphi_{\rm i}$ is from the i-th snapshot, and $i=1$ marks the earliest snapshot along this track, while $i=N$ marks the latest one. The jump described above manifests itself as a large difference of $\varphi$ between two adjacent snapshots. Therefore, we determine that a jump appears between $\varphi_{\rm i}$ and $\varphi_{\rm i+1}$ if $|\varphi_{\rm i+1}-\varphi_{i}|\geq 300^\circ$. $\varphi_{\rm i+1}-\varphi_{\rm i}<0$ means this jump is from $180^\circ$ to $-180^\circ$, and to recover a continuous $\varphi$ evolution, we make a transfer $\varphi_{\rm j}\to \varphi_{j}+360^\circ$ for all $j>i$. $\varphi_{\rm i+1}-\varphi_{i}>0$ means an opposite direction jump, and the corresponding transfer is $\varphi_{\rm j}\to \varphi_{\rm j}-360^\circ$ for all $j>i$. This correction is done sequentially from the earliest jump to the latest along the evolution track.

\subsubsection{Reasons for not using an even simpler frame}\label{sec:frame_choosing}
There is a spatial frame much simpler than that described in \S~\ref{sec:frame_construction}: choosing the plane spanned by the initial halo spin, $\boldsymbol{\lambda}_{\rm initial}$, and final halo spin, $\boldsymbol{\lambda}_{\rm final}$, as the xy-plane, the z-axis is parallel to $\boldsymbol{\lambda}_{\rm initial} \times \boldsymbol{\lambda}_{\rm final}$, the x-axis is parallel to $\boldsymbol{\lambda}_{\rm initial}$, and the y-axis is set such that the three axes form a right-handed Cartesian frame. A spherical coordinate system can be attached to this frame in a standard way (the same as described in \S~\ref{sec:frame_construction}), and then its two angular coordinates (polar angle $\tilde{\theta}$ and azimuthal angle $\tilde{\varphi}$) can be used to describe the direction of $\boldsymbol{\lambda}$.

However, this frame have two disadvantages. First, the direction of its z-axis depends on the angular offset between $\boldsymbol{\lambda}_{\rm initial}$ and $\boldsymbol{\lambda}_{\rm final}$. This axis is defined to be parallel to $\boldsymbol{\lambda}_{\rm initial} \times \boldsymbol{\lambda}_{\rm final}$. The direction of this vector product is determined through the right-hand rule, which requires to rotate $\boldsymbol{\lambda}_{\rm initial}$ to $\boldsymbol{\lambda}_{\rm final}$ through an angle smaller than $180^\circ$. If the angle between $\boldsymbol{\lambda}_{\rm initial}$ and $\boldsymbol{\lambda}$ gradually increase, but at the final moment it is still below $180^\circ$, then this frame represent this direction evolution as an angular increase, as expected. If the angle between $\boldsymbol{\lambda}_{\rm initial}$ and $\boldsymbol{\lambda}$ gradually increase and eventually excesses $180^\circ$, then the vector product rule would reverse the direction of z-axis, and therefore, the positive side of the xy-plane is changed, and the angle evolution in this frame is represented as an angular decrease. Because the redshift interval for spin direction evolution (from $z\sim 3.3$ to $z=0$) is large, the above second case does happen in the simulations we use. This makes halo samples a mixture of these two cases, and masks some statistical features of spin direction evolution. The frame described in \S~\ref{sec:frame_construction} largely avoid this problem, because there the vector product is only used on spin vectors from adjacent snapshots, of which the angular offsets are usually small.

Second, in this frame, by definition, $\boldsymbol{\lambda}$ lies exactly within the xy-plane at the initial and final moment. Correspondingly, at these two moments, $\tilde{\theta}=90^\circ$, though the evolution between them leads to other values of  $\tilde{\theta}$. This behaviour is not easy to incorporate into a simple stochastic model, as we will do in \S~\ref{sec:simple_model} for results derived with the frame described in \S~\ref{sec:frame_construction}.

\section{Results}\label{sec:results}
\subsection{Correlation between halo spin direction change and mass increase}\label{sec:j_m_relation}
\begin{figure*}
	\centering
	\includegraphics[width=1.0\textwidth]{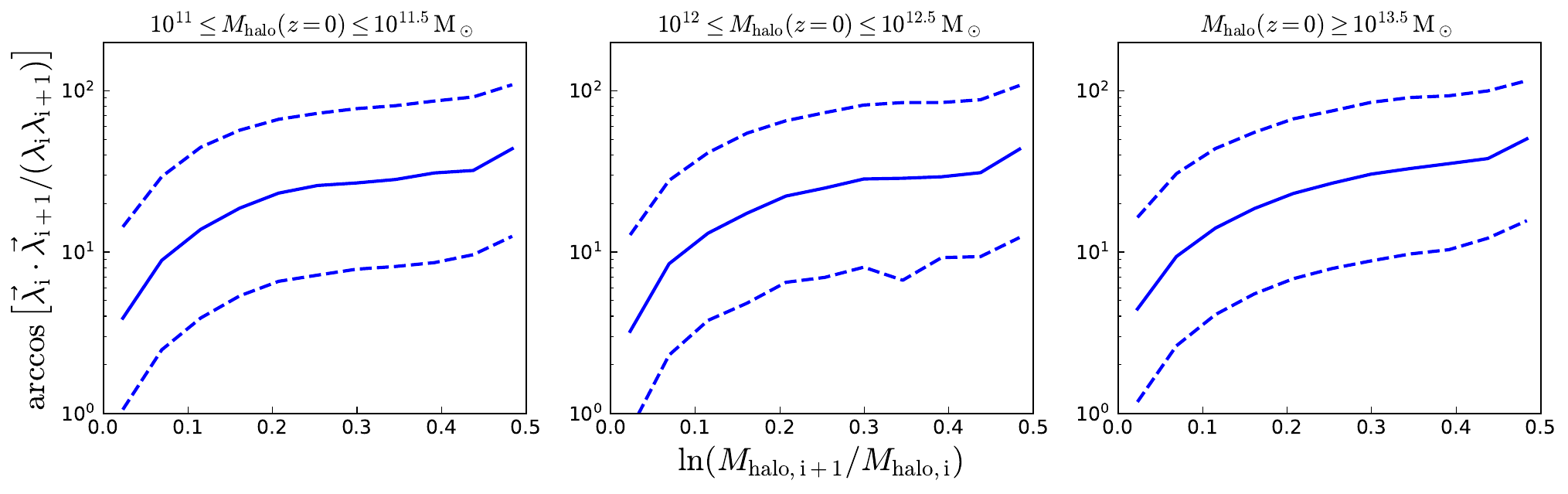}
	\caption{Halo spin direction changes between two adjacent snapshots correlate with halo mass changes within the same interval. Each panel is for a halo sample introduced in \S~\ref{sec:halo_sample}, with the $z=0$ halo mass range given at the top. In each panel, the solid line represents medians, while the dashed lines indicate 10-90 percentile ranges.}	
	\label{fig:angle_change_vs_mass_change}
\end{figure*}

Fig.~\ref{fig:angle_change_vs_mass_change} compares the spin direction changes between two adjacent snapshots and the halo mass increases within the same interval. Here the angle between spin vectors from two adjacent snapshots is adopted as a concise indicator for direction change strength. This angle can be derived through the dot product of these two vectors.

This figure shows a clear positive correlation between spin direction change and halo mass increase. A straightforward explanation for it would be that the spin direction evolution is mainly driven by the angular momentum accretion associated with the mass accretion. If, however, the dominant factors are something else, then to generate the correlation seen in Fig.~\ref{fig:angle_change_vs_mass_change}, these factors must correlate with halo mass accretion. Known factors other than angular momentum accretion, such as torques from surrounding structures and flyby objects, seems not have obvious correlations with halo mass accretion. Therefore, the previous straightforward explanation seems to be valid, at least approximately. Because of this, when we express further results of spin direction evolution, we choose $x\equiv\ln(M_{\rm halo}/M_{\rm halo,initial})$, where $M_{\rm halo,initial}$ is the initial halo mass, as the basic variable, instead of redshifts or physical time.

\subsection{Median evolution of spin direction}\label{sec:median_evolution}
\begin{figure*}
	\centering
	\includegraphics[width=1.0\textwidth]{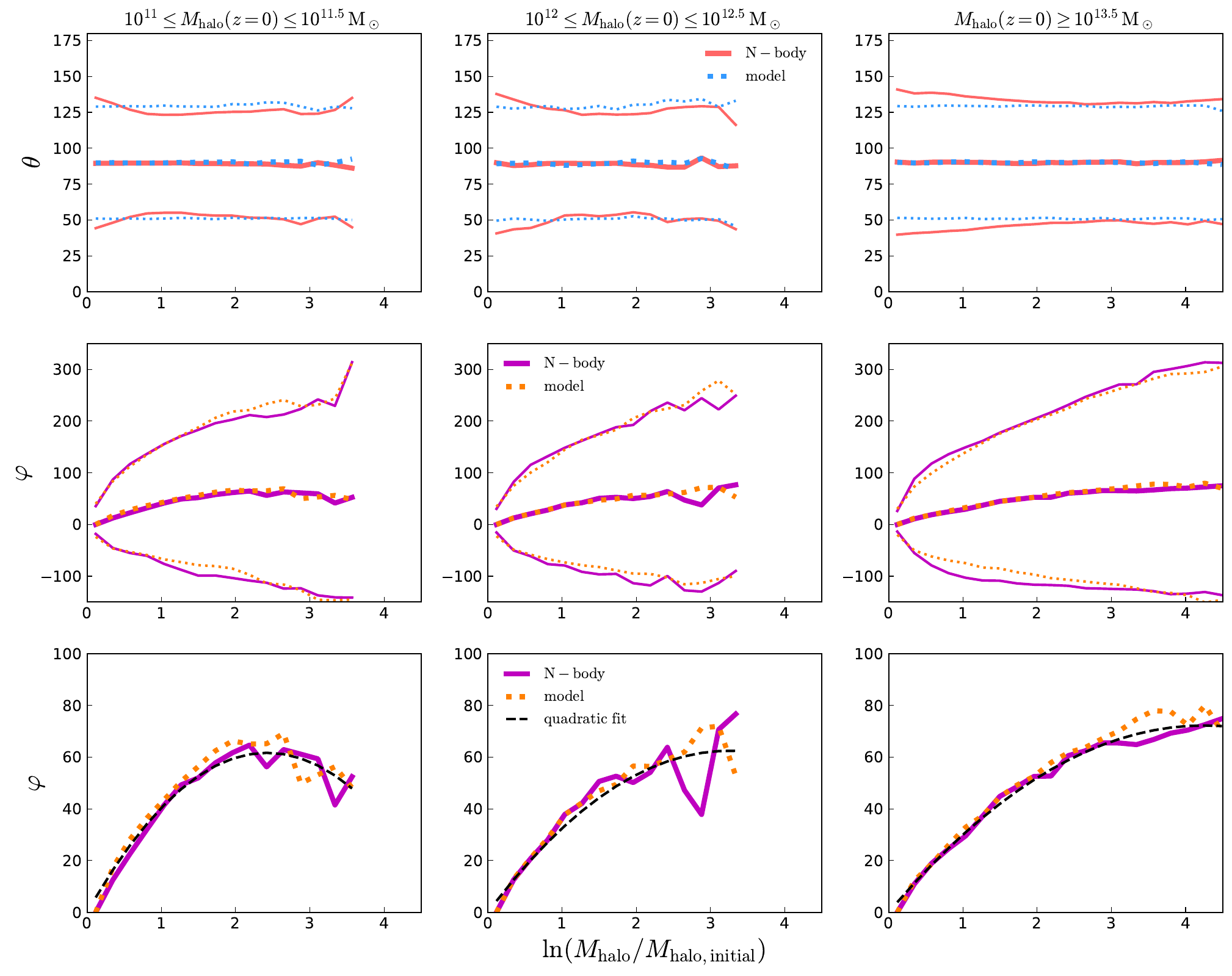}
	\caption{Median spin direction evolution. The angle $\theta$ and $\varphi$ for describing spin direction are defined in \S~\ref{sec:frame_construction}. Each column is for a halo sample, with corresponding $z=0$ halo mass range given at the top. In each panel, the thick lines indicate medians, while the thin lines of the same type indicate corresponding 10-90 percentile ranges. The third row zooms in the medians shown in the corresponding panels in the second row.}	
	\label{fig:angle_evolution}
\end{figure*}

Fig.~\ref{fig:angle_evolution} shows the evolution of $\theta$ and $\varphi$ against $x\equiv\ln(M_{\rm halo}/M_{\rm halo,initial})$. $\theta$ shows little evolution, with a stable median about $90^\circ$. This means that the spin vector, $\boldsymbol{\lambda}$, tends to oscillate around the average plane. 

The direction evolution within the average plane is very different. The medians of $\varphi$ tend to increase with $x$, except for $x>2.3$ in low and intermediate mass halo samples. Note that in these two samples, merger tree main branches stretching to $x>2.3$ are not very abundant, and therefore, the decrease may be at least partially caused by non-perfect statistics. The increase of $\varphi$ with $x$ indicates that there are coherent components in spin direction evolution, and this is consistent with the results in \citet{halo_spin_Contreras}. Note that the coherent evolution of $\varphi$ manifesting itself as an increase of $\varphi$ with $x$ is due to our choice of specific spatial frames for describing spin directions (described in detail in \S\ref{sec:frame_construction}). Reversing the direction of z-axes of these frames would lead the coherent evolution of $\varphi$ to appear as a decrease of $\varphi$ with $x$. If a fixed frame is used for all haloes, then this coherent evolution would be masked by the variety of halo orientations relative to this frame. For $\varphi$, the scatter around median seems to increase with $x$. This is also very different from the case for $\theta$, where the scatter around median is approximately constant against $x$.

\begin{figure*}
	\centering
	\includegraphics[width=1.0\textwidth]{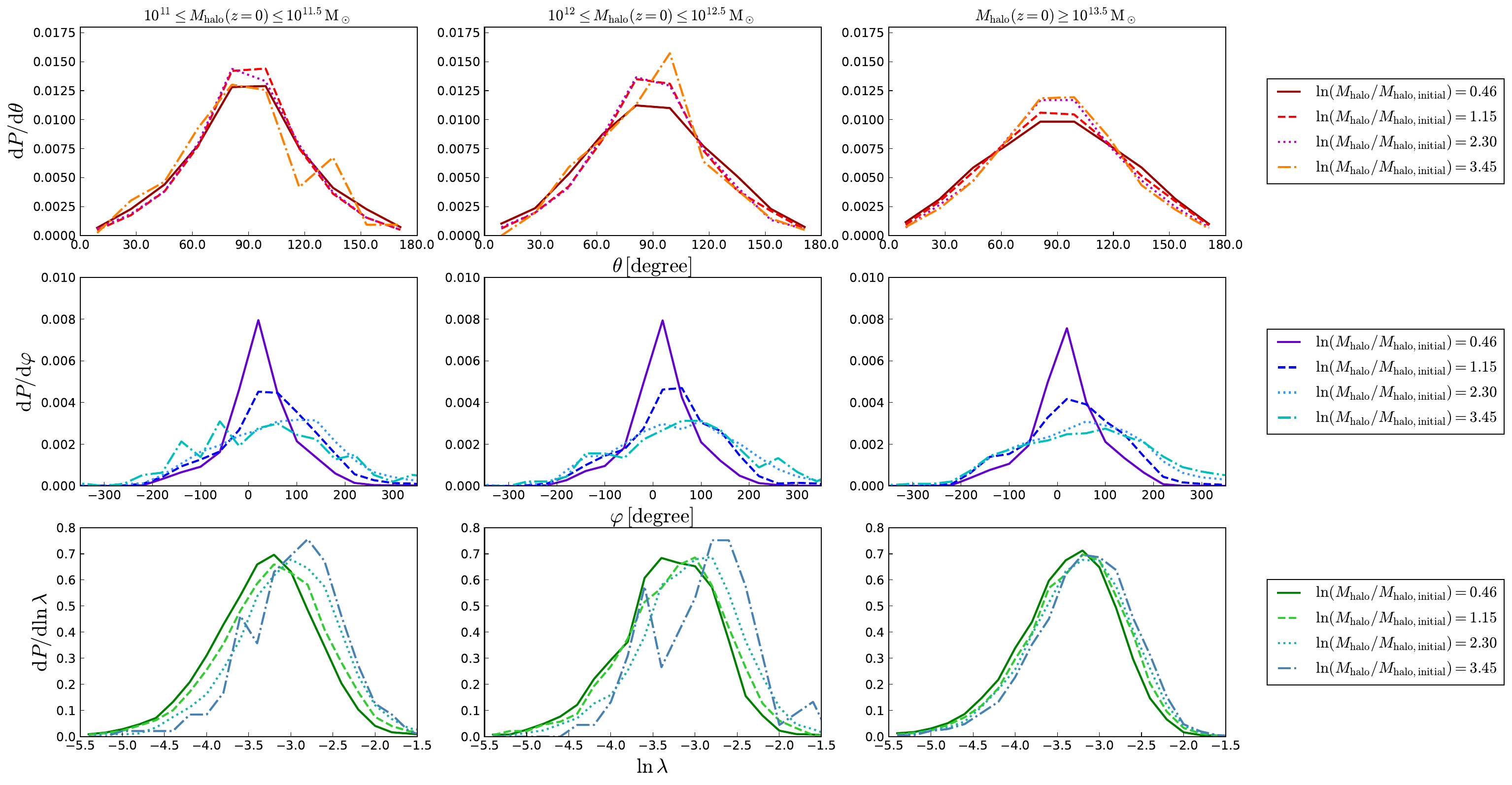}
	\caption{Probability distributions of $\theta$, $\varphi$ and $\ln\lambda$ at different values of $x$, where $x=\ln(M_{\rm halo}/M_{\rm halo,initial})$. The selected values of $x$ are given in the line key for each row. These selected $x$ correspond respectively to $\log_{10}(M_{\rm halo}/M_{\rm halo,initial})=0.2,\ 0.5,\ 1.0$ and $1.5$. Each row is for a halo sample, with corresponding $z=0$ halo mass range given at top.}	
	\label{fig:angle_distribution}
\end{figure*}

The evolution behaviours of $\theta$ and $\varphi$ are further investigated through comparing their probability distributions at different $x$. This is shown in Fig.~\ref{fig:angle_distribution}. As can be seen from this figure, the distributions of $\theta$ are approximately stable with $x$, but some minor evolutions also can be observed. These evolutions may stem from the factors other than angular momentum accretion that affect spin directions. They also may be caused by statistical biases left in the averaging process that derives the normal vector of the average plane (equation~\ref{eq:average_norm_v}). The peaks of $\varphi$ distributions move to larger $\varphi$ as $x$ increases, which again shows a coherent evolution of $\varphi$. These distributions become wider and wider with increasing $x$. This behaviour looks like a kind of diffusion.

Fig.~\ref{fig:angle_distribution} also shows the distributions of $\ln\lambda$ at different $x$. It seems that the peaks of these distributions gradually shift to higher $\ln\lambda$ with increasing $x$. For most of $x$ values shown in this figure, this shift is small, but it becomes quite obvious at $x=3.45$ for the low and intermediate mass samples. There are only about $3\%$ of main branches in these samples reaching such large values of $x$, and therefore, when consider the majority part of a sample, the evolution corresponding to this shift could be treated as a secondary feature, implying that the distribution of $\ln\lambda$ is approximately stable against $x$. The existence of this shift indicates that there are (secondary) correlations between halo spins and halo mass growth histories. This is consistent with previous works \citep[e.g.][]{lambda_evolution_Kim}.

The widely different behaviours of $\theta$ and $\varphi$ imply that the average plane introduced in spin direction measurement should correspond to some physical features in halo growth. There have been several works pointed out that haloes in filaments tend to have their spins perpendicular to filament spines when they are in active mass accretion phase [e.g.\ \citet{spin_filament_Borzyszkowski,spin_filament_Ganeshaiah1} and \citet{spin_filament_Ganeshaiah3}]. This trend limits major halo spin direction changes to the plane that is perpendicular to filament spines. \citet{spin_filament_Ganeshaiah3} also pointed out that haloes in walls tend to have spins lying within the wall plane. This trend limits major spin direction changes to the wall plane. These results provide possible explanations to the existence of an average plane observed in this work. Further investigations of this would be a task of future works.

\subsection{Autocorrelations of $\theta$, $\varphi$, and $\ln\lambda$}\label{sec:autocorrelations}
\begin{figure*}
	\centering
	\includegraphics[width=1.0\textwidth]{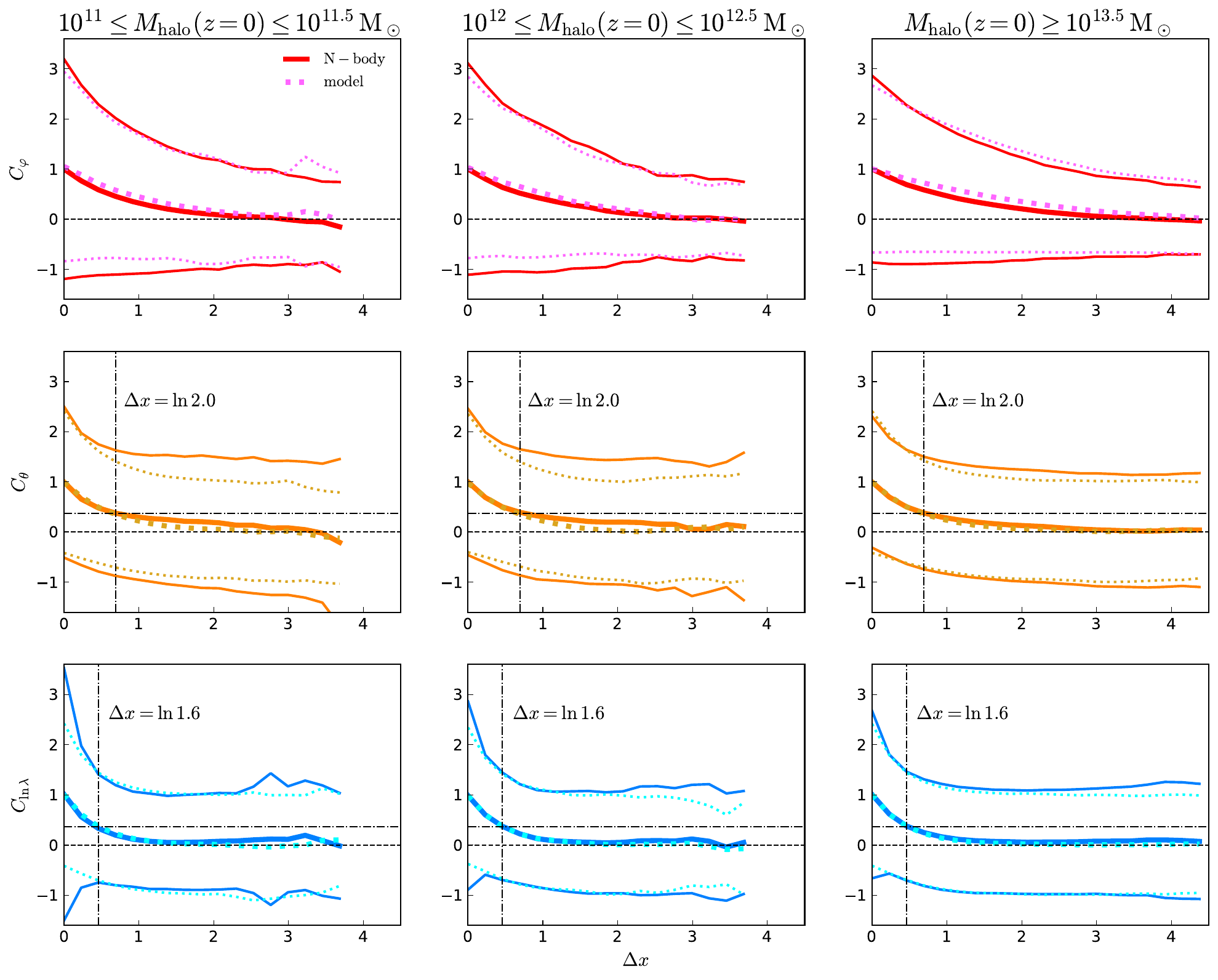}
	\caption{Autocorrelations of $\theta$, $\varphi$, and $\ln\lambda$. Each row is for a halo sample, with corresponding $z=0$ halo mass range given at top. In each panel, the thick lines indicate autocorrelations, while the thin lines of the same type indicate the standard deviations. The horizontal dashed line in each panel marks the constant $0$. In the lower two rows, the vertical and horizontal dotted-dashed lines mark the $\Delta x$ at which the autocorrelation measured from simulations drops to $1/e$. This parameter will be used in the stochastic model described in \S~\ref{sec:simple_model}.}	
	\label{fig:auto_correlation}
\end{figure*}

In this work, we calculate the autocorrelation of a quantity $A$ through
\begin{eqnarray}
C_{\rm A}(\Delta x) & = & \frac{1}{N_{\rm data}} \sum_{\rm i=1}^{N_{\rm branch}}\sum_{j=1}^{N_{\rm snap}(i)} 
\left(\frac{A(x_{\rm j}+\Delta x)-\bar{A}}{\sigma_{\rm A}}\times \right. \nonumber \\
 & & \left. \frac{A(x_{\rm j})-\bar{A}}{\sigma_{\rm A}}\right),
\label{eq:cal_auto_correlation}
\end{eqnarray}
where the first summation for all merger tree main branches in a sample, the second for all snapshots of a given main branch, $\bar{A}$ and $\sigma_{\rm A}$ are respectively the mean and standard deviation of $A$ among this sample, and $N_{\rm data}$ is the total number of pairs that appear in the summations.

The autocorrelations of $\theta$, $\varphi$, and $\ln\lambda$ as functions of $\Delta x$ are shown in Fig.~\ref{fig:auto_correlation}. All three kind of autocorrelations gradually drop to zero when $\Delta x$ is large. This means that halo growth gradually erases features of the original halo spin vector. Among the three quantities, $\varphi$ shows the strongest correlation, consistent with the fact that it is the only quantity showing coherent evolution.

\subsection{cross-correlations between $\theta$, $\varphi$, and $\ln\lambda$}\label{sec:cross_correlations}
In this work, we calculate the cross-correlation of quantities $A$ and $B$ through
\begin{eqnarray}
C_{\rm AB}(\Delta x) & = & \frac{1}{N_{\rm data}} \sum_{\rm i=1}^{N_{\rm branch}}\sum_{j=1}^{N_{\rm snap}(i)} 
\left(\frac{A(x_{\rm j}+\Delta x)-\bar{A}}{\sigma_{\rm A}}\right.\times \nonumber \\
& & \left. \frac{B(x_{\rm j})-\bar{B}}{\sigma_{\rm B}}\right).
\label{eq:cal_cross_correlation}
\end{eqnarray}
Similar to equation~(\ref{eq:cal_auto_correlation}) for autocorrelations, here the first summation for all merger tree main branches in a sample, the second for all snapshots of a given main branch, and $N_{\rm data}$ is the total number of pairs that appear in the summations. $\bar{A}$ and $\bar{B}$ are respectively the means of $A$ and $B$ among this sample, while $\sigma_{\rm A}$ and $\sigma_{\rm B}$ are corresponding standard deviations.

\begin{figure*}
	\centering
	\includegraphics[width=1.0\textwidth]{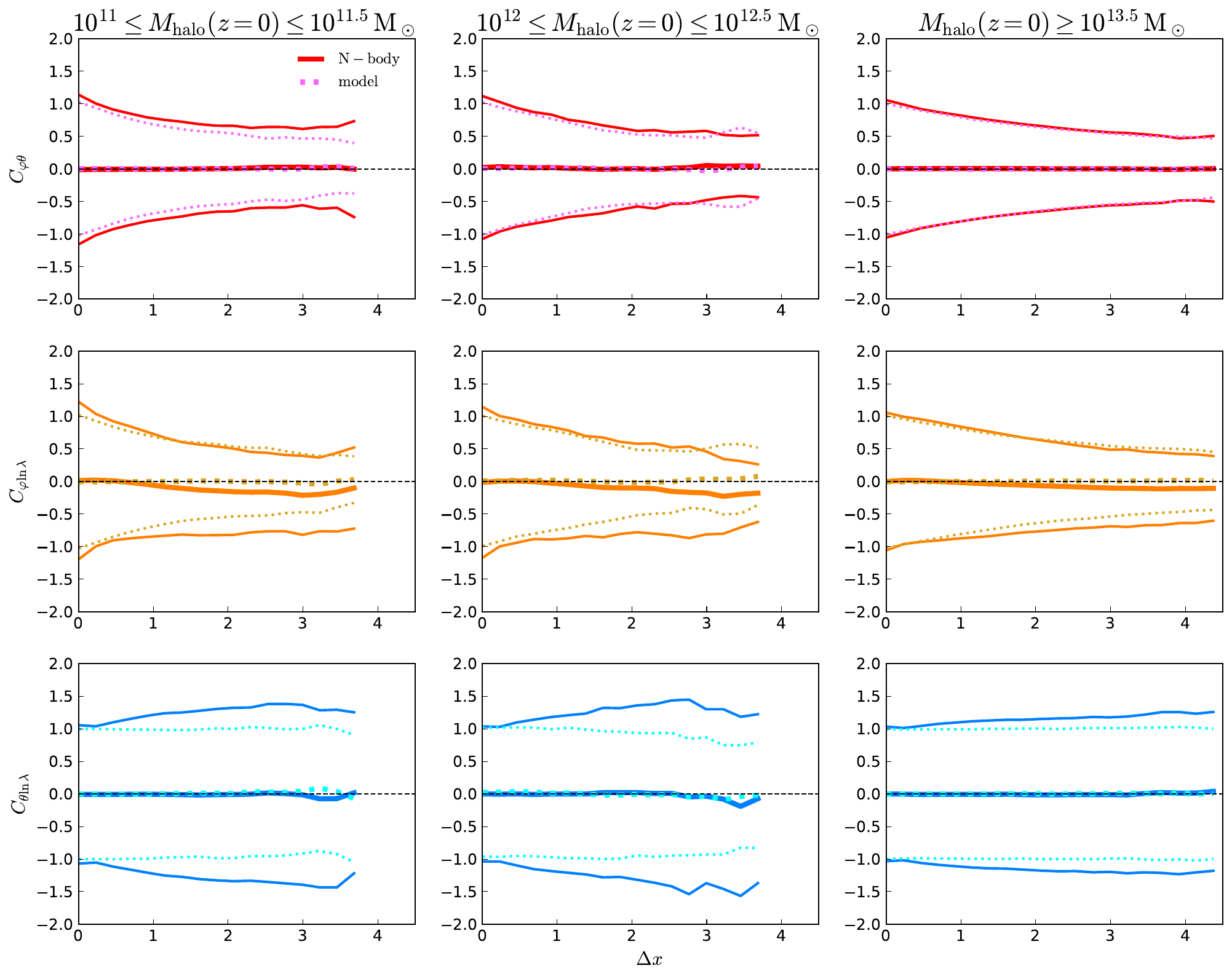}
	\caption{cross-correlations of $\theta$, $\varphi$, and $\ln\lambda$. Each row is for a halo sample, with corresponding $z=0$ halo mass range given at top. In each panel, the thick lines indicate cross-correlations, while the thin lines of the same type indicate the standard deviations. The horizontal dashed line in each panel marks the constant $0$.}	
	\label{fig:cross_correlation}
\end{figure*}

The results are shown in Fig.~\ref{fig:cross_correlation}. It seems that all cross-correlations are close to zero for all values of $\Delta x$. This indicates that the correlations between $\theta$, $\varphi$ and spin magnitude are weak, and in the lowest order approximation, their evolutions can be treated as independent.

\subsection{Simple stochastic model}\label{sec:simple_model}
\subsubsection{Model description} \label{sec:simple_model_description}
Based on results shown in previous sections, we view the evolution of halo spin vectors as a random walk with $x=\ln(M_{\rm halo}/M_{\rm halo,initial})$ its basic variable. According to the cross-correlations shown in Fig.~\ref{fig:cross_correlation}, we assume the evolutions of $\theta$, $\varphi$ and $\ln \lambda$ are independent. This assumption leads to significant simplification, because it allows us to treat their evolutions separately.

We first consider the evolution of $\ln \lambda$. It is well known that the probability distribution of this quantity is approximately gaussian and very stable against redshifts and halo masses \citep[e.g.][]{lambda_evolution_Kim}. For the conciseness of latter expressions, below we consider the evolution of $y_{\lambda}\equiv(\ln\lambda-\overline{\ln\lambda})/\sigma_{\lambda}$, where $\overline{\ln\lambda}$ and $\sigma_{\lambda}$ are respectively the mean and standard deviation of the stable distribution of $\ln\lambda$. If $\ln\lambda$ has a stable gaussian distribution, then $y_{\lambda}$ has a gaussian distribution with zero mean and unit scatter as its stable distribution.

The simplest Langevin equation for $y_{\lambda}$ that ensures the above feature has the form:
\begin{equation}
\frac{d y_{\lambda}}{dx}=-\frac{y_{\lambda}}{\tau_{\lambda}}+\sqrt{\frac{1}{\tau_{\lambda}}}\xi_{\lambda}(x),
\label{eq:lambda_L_eqn}
\end{equation}
where $\tau_{\lambda}$ is a parameter that determines the speed of relaxation towards the stable distribution [this becomes obvious in equation~(\ref{eq:P_lambda})], and $\xi_{\lambda}(x)$ is a gaussian white noise.

The corresponding Fokker-Planck equation is
\begin{equation}
\frac{\partial P}{\partial x}=\frac{1}{\tau_{\lambda}}\left( \frac{\partial}{\partial y_{\lambda}}(y_{\lambda}P)+\frac{\partial^2}{\partial y_{\lambda}^2}P\right),
\label{eq:lambda_FP_eqn}
\end{equation}
where $P\equiv P(y_{\lambda},x)$ is the probability distribution of $y_{\lambda}$ at $x$.

If $y_{\lambda}=y_0$ at $x=x_0$, then the above equation gives the conditional probability distribution of $y_{\lambda}$ for $x>x_0$:
\begin{eqnarray}
P(y_{\lambda},x|y_0,x_0) & = & \frac{1}{\sqrt{2\pi(1-e^{-\frac{2\Delta x}{\tau_\lambda}})}} \times \nonumber \\
 & & \exp\left[-\frac{(y_{\lambda}-y_0e^{-\Delta x/\tau_{\lambda}})^2}{2(1-e^{-2\Delta x/\tau_{\lambda}})}  \right],
\label{eq:P_lambda}
\end{eqnarray}
where $\Delta x=x-x_0$. 

Given $P(y_{\lambda},x|y_0,x_0)$ [denoted below as $\tilde{P}(y_{\lambda})$ for conciseness], the autocorrelation, $C(\Delta x)$, of evolution tracks with $y_0$ randomly picked from $y_{\lambda}$'s stable distribution, $P(y_0)$, can be calculated as:
\begin{eqnarray}
C(\Delta x) & = & \int\int y_{\lambda}(\Delta x+x_0)y_0\tilde{P}(y_{\lambda})P(y_0)dy_{\lambda}dy_0 \nonumber \\
 & = & e^{-\Delta x/\tau_{\lambda}},
\end{eqnarray}
which gives that $C(\tau_{\lambda})=1/e$. Given the autocorrelations measured from N-body simulations (Fig.~\ref{fig:auto_correlation}), $\tau_{\lambda}$ can be determined. The simulations used in this work give $\tau_{\lambda}=\ln 1.6$.

With $\tau_{\lambda}$ given, the conditional distribution of $y_{\lambda}$ with initial value $y_0$ is fully determined. It can be transferred to distributions of $\ln\lambda$ straightforwardly if $\overline{\ln\lambda}$ and $\sigma_{\lambda}$ are known. In this work, we estimate them through the mean and standard deviation of $\ln\lambda$ of haloes in all three samples. The estimations of $\overline{\ln\lambda}$ and $\sigma_{\lambda}$ also determine the stable distribution of $\ln\lambda$.

Now we consider the evolution of $\theta$. Its behaviour is similar to $\ln \lambda$. Assume the stable distribution of $\theta$ is approximately gaussian, and $\bar{\theta}$ and $\sigma_{\theta}$ are respectively its mean and standard deviation. Then through procedures entirely similar to those for $\ln\lambda$, we can derive the conditional probability distribution for $y_{\theta}\equiv(\theta-\bar{\theta})/\sigma_{\theta}$:
\begin{eqnarray}
P(y_{\theta},x|y_0,x_0) & = & \frac{1}{\sqrt{2\pi(1-e^{-\frac{2\Delta x}{\tau_\theta}})}} \times \nonumber \\
 & & \exp\left[-\frac{(y_{\theta}-y_0e^{-\Delta x/\tau_{\theta}})^2}{2(1-e^{-2\Delta x/\tau_{\theta}})}  \right].
\label{eq:P_theta}
\end{eqnarray}
$\tau_{\theta}$ can also be determined through the autocorrelations measured from N-body simulations. The simulations used in this work give $\tau_{\theta}=\ln 2$. The conditional distribution of $y_{\theta}$ can be transferred to that of $\theta$ given $\bar{\theta}$ and $\sigma_\theta$. Similar to the case for $\ln\lambda$, here we estimate them as the mean and standard deviation of $\theta$ of haloes in all three samples. These two estimations also fix the stable distribution of $\theta$.

Finally we consider the evolution of $\varphi$. Results in previous sections indicate that two major components in its evolution are coherent changing and diffusion. Therefore, we write the Langevin equation for $\varphi$ as
\begin{equation}
\frac{d\varphi}{dx}=A(x)+B\xi_{\varphi}(x),
\label{eq:phi_L_eqn}
\end{equation}
where $\xi_{\varphi}(x)$ is a gaussian white noise, and we assume $B$ is a constant for simplicity. The first term on the right-hand side of equation~(\ref{eq:phi_L_eqn}) represents $\varphi$'s coherent evolution, while the second term generates diffusion. Note that a white noise is uncorrelated along $x$, leading to a Markov random walk. This, however, does not contradict with \citet{halo_spin_Contreras}, because they mean spin direction evolutions are partially coherent and therefore have long range time correlations, while here we assume the random fluctuations around the coherent trend are timely uncorrelated.

The Fokker-Planck equation corresponding to equation~(\ref{eq:phi_L_eqn}) is:
\begin{equation}
\frac{\partial P}{\partial x}=-\frac{\partial}{\partial \varphi}\left[A(x)P\right]+B^2\frac{\partial^2}{\partial \varphi^2}P,
\label{eq:phi_FP_eqn}
\end{equation} 
where $P=P(\varphi,x)$ is the probability distribution of $\varphi$ at $x$. 

Then for $\varphi=\varphi_0$ at $x=x_0$, equation~(\ref{eq:phi_FP_eqn}) gives the conditional distribution of $\varphi$ for $x>x_0$ as:
\begin{eqnarray}
P(\varphi,x|\varphi_0,x_0) & = & \frac{1}{\sqrt{2\pi(k_\sigma\Delta x)}} \times \nonumber \\
& & \exp\left\{ -\frac{[\varphi-(\varphi_0+\alpha(\Delta x))]^2}{2(k_\sigma\Delta x)} \right\},
\label{eq:P_phi}
\end{eqnarray}
where $\Delta x=x-x_0$, $k_\sigma=2B^2$ and
\begin{equation}
\alpha(\Delta x)=\int_{x_0}^{x_0+\Delta x} A(\xi){\rm d}\xi.
\label{eq:define_alpha}
\end{equation}

The $\varphi$ evolution tracks in simulations all start from $\varphi_0=0$ at $x_0=0$. Plugging this into equation~(\ref{eq:P_phi}) one has
\begin{equation}
P(\varphi,x)=\frac{1}{\sqrt{2\pi(k_\sigma x)}}\exp\left\{ -\frac{[\varphi-\alpha(x)]^2}{2(k_\sigma x)} \right\}.
\label{eq:P_phi1}
\end{equation}
We found that quadratic functions provide good fits to the coherent median evolution of $\varphi$ in the N-body simulations (as can be seen from the bottom row of Fig.~\ref{fig:angle_evolution}), and therefore, in this model, we assume $\alpha(x)$ to also have a quadratic form:
\begin{equation}
\alpha(x)=\alpha_1 x^2+\alpha_2 x,
\label{eq:specific_alpha}
\end{equation}
where the two coefficients $\alpha_1$ and $\alpha_2$ are determined through fitting the simulation results. We measure in the simulations the variances of $\varphi$ at different $x$, which is denoted as $\sigma^2_{\varphi}(x)$, and then calculate $\sigma^2_{\varphi}(x)/x$. It is very stable against $x$, and we average it among all $x$ and all halo samples to estimate $k_{\sigma}$. This gives $k_{\sigma}=6718.5$. With $k_{\sigma}$ and $\alpha(x)$ given, the distributions of $\varphi$ are fixed.

The values of parameters in distributions of $\ln\lambda$, $\theta$ and $\varphi$ are summarized in Table~\ref{tab:model_parameters}.

\begin{table*}
	\centering
	\begin{tabular}{ccc|ccc|ccccccc}
		\hline
		\multicolumn{3}{c|}{$\ln\lambda$} & \multicolumn{3}{c|}{$\theta$} & & & \multicolumn{2}{c}{$\varphi$} & & & \\
		\hline
		\multicolumn{2}{c|}{stable} & conditional & \multicolumn{2}{c|}{stable} & conditional & & & \multicolumn{2}{c}{conditional} & & & \\
		\multicolumn{2}{c|}{distribution} & distribution & \multicolumn{2}{c|}{distribution} & distribution & & & \multicolumn{2}{c}{distribution} & & & \\
		\hline
		\multirow{2}{*}{$\overline{\ln\lambda}$} & \multirow{2}{*}{$\sigma_\lambda$} & \multirow{2}{*}{$\tau_\lambda$} & \multirow{2}{*}{$\bar{\theta}$} & \multirow{2}{*}{$\sigma_\theta$} & \multirow{2}{*}{$\tau_\theta$} & \multirow{2}{*}{$k_\sigma$} & \multicolumn{2}{c}{low-mass} & \multicolumn{2}{c}{intermediate} & \multicolumn{2}{c}{massive} \\
		\cline{8-13}
		& & & & & & & $\alpha_1$ & $\alpha_2$ & $\alpha_1$ & $\alpha_2$ & $\alpha_1$ & $\alpha_2$ \\
		\hline
		$-3.30$ & $0.64$ & $\ln 1.6$ & $90^\circ$ & $30.6^\circ$ & $\ln 2$ & $6718.5$ & $-10.5$ & $50.9$ & $-5.9$ & $38.4$ & $-4.0$ & $34.0$ \\
		\hline
	\end{tabular}
\caption{Values of parameters in distributions of $\ln\lambda$, $\theta$ and $\varphi$ used in the simple spin evolution model. These values are estimated from the N-body simulations used in this work. Here `low mass', `intermediate' and `high mass' correspond to the three halo samples used in this work, which are selected based on $z=0$ halo masses. The corresponding mass ranges are respectively $[10^{11}\,{\rm M}_\odot, 10^{11.5}\,{\rm M}_\odot]$, $[10^{12}\,{\rm M}_\odot, 10^{12.5}\,{\rm M}_\odot]$ and $[10^{13.5}\,{\rm M}_\odot, +\infty)$.}
\label{tab:model_parameters}
\end{table*}

For the halo at the tip of a given halo merger tree branch, set $\varphi=0$ and randomly pick $\ln\lambda$ and $\theta$ from corresponding stable distributions, and then its spin vector is fully determined. $\ln\lambda$ and $\theta$ of its descendants can be generated through conditional distributions given in equation~(\ref{eq:P_lambda}) and (\ref{eq:P_theta}). $\varphi$ of its descendants can be generated through using either equation~(\ref{eq:P_phi}) or (\ref{eq:P_phi1}). In this way, a spin vector is assigned to each halo in this branch. In a full merger tree with multiple branches, a descendant halo is assigned a spin along the branch containing its major progenitor (the most massive progenitor or the one containing its most bounded particles).

We generate model spins for main branches extracted from the N-body simulations used in this work, and derive statistical results in the same way as for N-body spin vectors. These model results are compared with corresponding simulation results, and are shown in Fig.~\ref{fig:angle_evolution}, Fig.~\ref{fig:auto_correlation} and Fig.~\ref{fig:cross_correlation}. 

In general, the model results are in good agreement with those of simulations, but there are some defects. The model seems not to reproduce the weak correlation between $\varphi$ and $\ln\lambda$ seen in the middle row of Fig.~\ref{fig:cross_correlation}. This is because all joint correlations between $\theta$, $\varphi$ and $\ln \lambda$ are ignored. This model also tends to slightly underestimate the standard deviations of $C_{\theta}$ (Fig.~\ref{fig:auto_correlation}), $C_{\varphi\ln\lambda}$ and $C_{\theta\ln\lambda}$ (Fig.~\ref{fig:cross_correlation}).

\subsubsection{A simple comparison with \citet{Benson2020}} \label{sec:model_comp}

We now compare the results from our spin evolution model with those from the \citet{Benson2020} model. The \citeauthor{Benson2020} model is a development of that of \citet{lambda_distri_Vitvistska}. It treats the halo evolution as a sequence of merger events, and calculates how the angular momentum changes in each halo merger event by statistically assigning orbital parameters to the haloes involved. The \citeauthor{Benson2020} model includes quite a large number of adjustable parameters, which are constrained by fitting to the distribution of spin parameters of relaxed halos at $z=0$ measured from an N-body simulation. \citeauthor{Benson2020} make several further comparisons between the predictions of their model and N-body simulations. Below, we make the same comparisons between the predictions of our model and N-body simulations, and also compare with the predictions of the \citeauthor{Benson2020} model. We derive N-body simulation measurements as described in \citeauthor{Benson2020}, except that we use Dhalos instead of FOF groups to define the halos. These differences do not change the conclusions in \citeauthor{Benson2020}. The halo sample used in \citeauthor{Benson2020} approximately corresponds to the sample of massive haloes in this work. Here we extend the comparison between N-body measurements and model predictions to also include samples of less massive haloes.

\begin{figure*}
	\centering
	\includegraphics[width=1.0\textwidth]{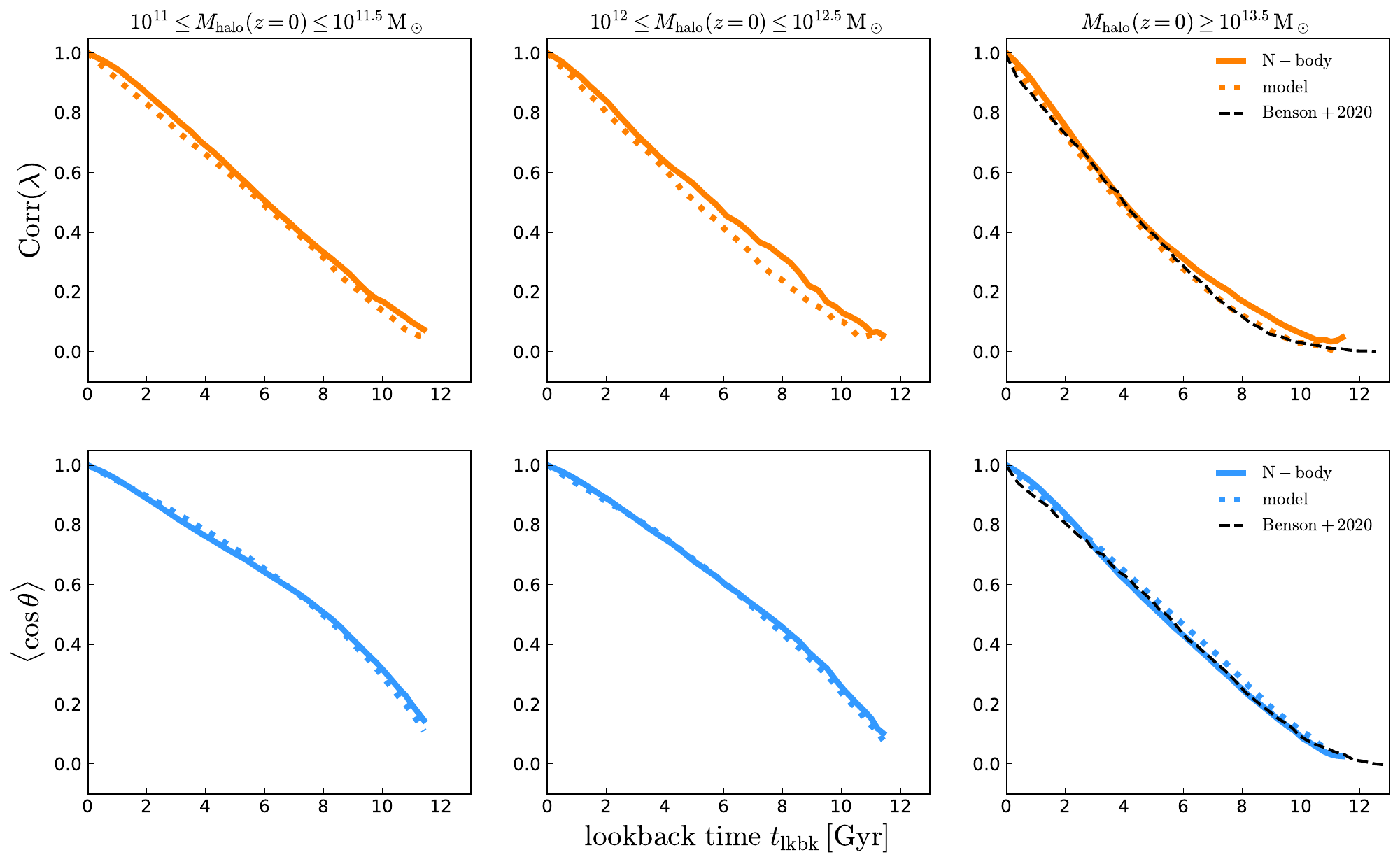}
	\caption{{\it Top row}: the correlation function of spin parameter $\lambda$ at $z=0$ and at look-back time $t_{\rm lkbk}$. Each panel is for a different $z=0$ halo mass range, as shown at the top. {\it Bottom row}: The angle between the spin vector at $z=0$ and that at look-back time $t_{\rm lkbk}$, for the different halo mass ranges.}	
	\label{fig:Benson2020_correlation}
\end{figure*}

The top row of Fig.~\ref{fig:Benson2020_correlation} shows the correlation function between $\lambda$ at $z=0$ and that at lookback time $t_{\rm lkbk}$. The \citet{Benson2020} model reproduces this correlation well in the halo mass range that it considered. In this mass range, the prediction of our model is also in good agreement with N-body measurements. Furthermore, our model can also reproduce these correlations for less massive haloes.

The bottom row of Fig.~\ref{fig:Benson2020_correlation} shows the median angle between the spin vector at $z=0$ and that at lookback time $t_{\rm lkbk}$. Again, the \citet{Benson2020} model reproduces this well in the halo mass range that it considered (massive haloes), while our model can reproduce this for lower mass haloes as well as for massive haloes. Note that this angle does not provide full information of how spin direction changes in 3D space. The results in section~\ref{sec:median_evolution} provide a more complete description of the evolution in spin direction.

\begin{figure*}
	\centering
	\includegraphics[width=1.0\textwidth]{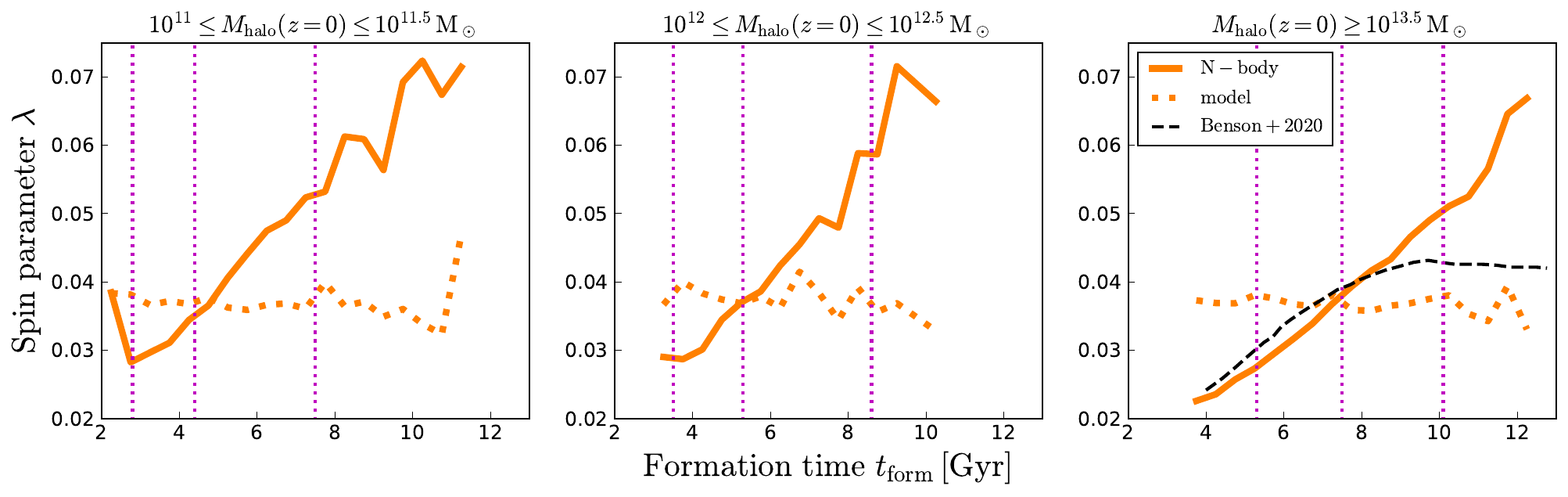}
	\caption{Median spin parameter of $z=0$ haloes as a function of halo formation time $t_{\rm form}$. Here the halo formation time is defined as the cosmic time at which the haloes along the merger tree main branch first exceed half of the $z=0$ halo mass. Each panel is for a different $z=0$ halo mass range, with the corresponding values given on the top. In each panel, from left to right, the three vertical dotted lines respectively indicate 10, 50 and 90 percentiles of halo formation times of the corresponding halo sample.}		
\label{fig:Benson2020_t_form_correlation}
\end{figure*}

Fig.~\ref{fig:Benson2020_t_form_correlation} shows the relation between the median $\lambda$ for $z=0$ halos and halo formation time, $t_{\rm form}$. Here the formation time is defined as the time at which the main progenitor halo first reaches $50\%$ of the mass of its $z=0$ descendant halo. The measurements from N-body simulations show a clear positive correlation, while our model predictions are essentially flat, i.e.\ no correlation. The flatness is caused by the assumption in our random walk model that the spin evolution only depends on the change in halo mass, independent of when this mass change happens. Whatever the value of the formation time is, the mass increase from this time to the present is the same, i.e.\ half the present mass, and therefore our model gives statistically the same random walk to haloes with the same $z=0$ mass but different formation times.

\begin{figure*}
	\centering
	\includegraphics[width=1.0\textwidth]{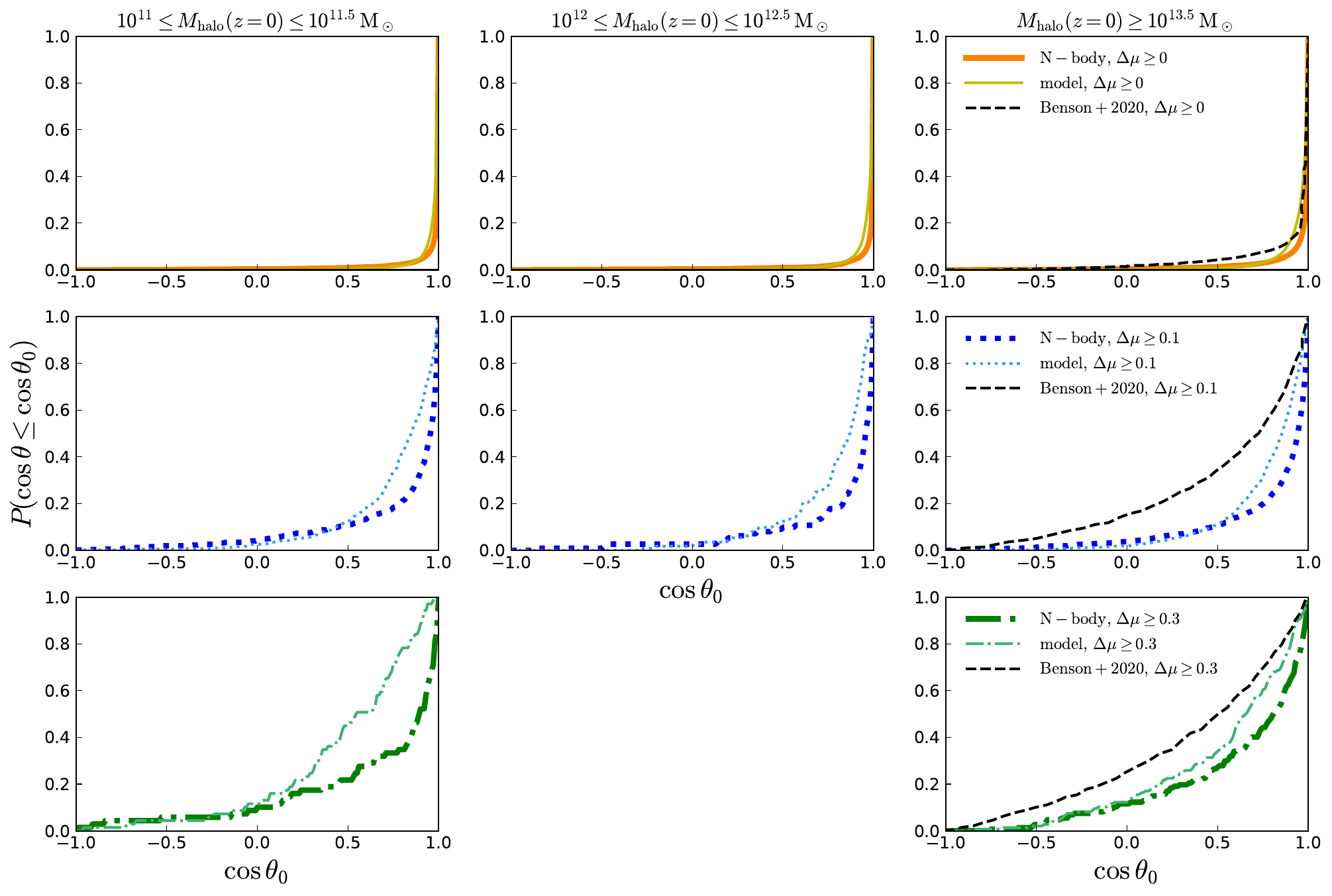}
	\caption{Cumulative distributions of the angle between spin vector at $z=0$ and that $0.54\,{\rm Gyr}$ (two snapshots) ago. Each column is for a different $z=0$ halo mass range, with the values given along the top. The top row includes all haloes, while the middle row only includes haloes with mass increases in the last $0.54\,{\rm Gyr}$ larger than $10\%$ of their final masses, and the bottom row only includes haloes with mass increases larger than $30\%$ of their final mass. The middle panel of the bottom row is omitted, because, for that sample, there are too few haloes to derive statistically reliable results.}		
\label{fig:Benson2020_spin_flips}
\end{figure*}

Fig.~\ref{fig:Benson2020_spin_flips} shows the cumulative distributions of spin-flip angles. Following \citet{Benson2020}, these are defined as the angle between spin vector at $z=0$ and that $0.54\,{\rm Gyr}$ (two snapshots) ago. The top row of this figure, which includes all haloes in each mass range, shows that both the model in \citeauthor{Benson2020} and our model can reproduce the distribution for the massive halo sample. It further shows that our model can also reproduce the distributions for the two lower halo mass ranges.

\citet{Benson2020} then investigated the cumulative distributions of spin-flip angles for haloes having large mass increases within this $0.54\,{\rm Gyr}$ interval. Here the strength of the mass increases is described by $\Delta\mu$, which is defined as the ratio of mass increase to the $z=0$ halo mass. \citeauthor{Benson2020} considered two sub-samples of the massive halo sample with respectively $\Delta\mu\geq 0.1$ and $\Delta\mu\geq 0.3$. In general, larger mass increases are more likely to have large spin-flip angles. The model in \citeauthor{Benson2020} can qualitatively reproduce this trend, but quantitatively overestimates the fraction of large flip angles. The predictions of our model seem to be in better agreement with the N-body measurements than the model in \citeauthor{Benson2020}, but it still overestimates the fraction of intermediate spin flips ($\cos \theta_0\sim 0.5$), as can be seen from the middle and bottom rows of Fig.~\ref{fig:Benson2020_spin_flips}. Note that we do not show the distribution of haloes with $\Delta\mu\geq 0.3$ for the sample with $z=0$ halo masses in $[10^{12}\,{\rm M}_\odot,10^{12.5}\,{\rm M}_\odot]$, because this sub-sample contains only $9$ haloes, and statistically reliable results cannot be derived from it.

\section{Summary}\label{sec:summary}
In this work we consider the evolution of the halo spin vector along the main branches of dark matter halo merger trees extracted from the Millennium-I and Millennium-II N-body simulations. Here the main branches start from the base node at $z=0$ and contain the most massive progenitor halo in each branch to higher redshift. We selected three halo samples according to their $z=0$ masses. The corresponding base node mass ranges are respectively $[10^{11}\,{\rm M}_\odot,10^{11.5}\,{\rm M}_\odot]$, $[10^{12}\,{\rm M}_\odot,10^{12.5}\,{\rm M}_\odot]$ and $[10^{13.5}\,{\rm M}_\odot,+\infty)$. The first two samples are taken from the Millennium-II simulation, because of its high mass resolution, while the third sample is from the Millennium-I simulation, for its large simulation volume. We limited the measurement of halo spins to well resolved haloes (resolved with at least $300$ particles), and this gives spin vector evolution between $z=0$ and $z\sim 3.3$.

We found that changes in halo spin direction correlate strongly with changes in halo mass.
Furthermore, for each main branch of the halo merger tree, there seems to be a characteristic plane for evolution of the spin direction. In the direction perpendicular to this plane, halo spins tend to oscillate around this plane. The distribution of angles between spin vectors and this plane tends to be approximately stable. Within this plane, halo spins show a coherent direction change with time, as well as a diffusion in directions around this average trend. The correlation between the evolution in directions within and perpendicular to the plane is weak, and the correlation of these directions with the magnitude of the spin also seems to be weak. Previously \citet{halo_spin_Contreras} reported that the halo spin direction evolution is time-correlated and therefore inferred that this evolution is partially coherent, while in this work we analyse this effect in more detail. This characteristic plane could be related to the geometry of the surrounding large-scale structure, but the physical mechanism behind the coherent evolution in spin direction is currently not clear.

We also constructed a simple stochastic model for spin vector evolution, in which this evolution is viewed as a random walk taking logarithmic halo mass as its basic variable. This model can approximately reproduce the statistical results of spin evolution measured from the two N-body simulations used in this work, including the median direction evolution, the scatter around this median evolution, and various auto-correlations and cross-correlations. However, this model does not reproduce the correlation between spin magnitudes and halo formation times seen in simulation results. Future work may improve on this. This simple model for the evolution of the halo spin vector could be an alternative choice for simplified (semi-analytical) galaxy formation models, either when the N-body simulations for halo merger tree construction do not have high enough resolution to allow a large fraction of haloes in merger trees to have reliable spins measured directly, or when Monte Carlo merger trees are used.

\section*{Acknowledgements}
JH thank Dr. John Helly for his patient and detailed instruction of Dhalo merger tree usage, and thank Dr. Jian Fu for some helpful discussions. 
This work was supported by the Natural Science Foundation of Shanghai [20PJ1412300] and National Natural Science Foundation of China [12003015]. CGL acknowledges support from STFC (ST/X001075/1).
This work used the DiRAC@Durham facility managed by the Institute for Computational Cosmology on behalf of the STFC DiRAC HPC Facility (www.dirac.ac.uk). The equipment was funded by BEIS capital funding via STFC capital grants ST/K00042X/1, ST/P002293/1, ST/R002371/1 and ST/S002502/1, Durham University and STFC operations grant ST/R000832/1. DiRAC is part of the National e-Infrastructure.

\section*{Data availability}
The data underlying this article will be shared on reasonable request to the corresponding author.

\bibliographystyle{mn2e}

\bibliography{paper}

\appendix
\section{Pseudo-code of the spin vector evolution model}
In this appendix we provide a pseudo-code of our spin vector evolution model. Before apply this model, one needs to determine the values of its parameters according to the base node mass ($z=0$ halo mass). These values can be found in Table~\ref{tab:model_parameters}. In this pseudo-code, we assume a merger tree can be walked through in a specific way, which starts from a halo at the highest redshift level, and moves to a lower redshift level only if all haloes in the current level have been visited. We use the following conventions: 

\begin{description}
	\item [$\leftarrow$:] implies assignment;
	\item [$tree$:] the merger tree being processed;
	\item [$halo$:] the current halo being processed;
	\item [$child\_halo$:] the main progenitor halo of the current halo;
	\item [$\cdot firstHalo$:] an operator which returns the first halo for a walk of a merger tree;
	\item [$\cdot child$:] an operator which returns the main progenitor of the current halo, or $null$ if no progenitor;
	\item [$\cdot next$:] an operator which returns the next halo in a walk of a merger tree, or $null$ if no more haloes remain;
	\item [$\cdot mass$:] an operator which returns the mass of the current halo;
	\item [$\cdot\theta$:] an operator which returns $\theta$ of the spin vector of the current halo;
	\item [$\cdot x$:] an operator which returns $x$ of the current halo;
	\item [$\cdot sample$:] an operator which samples from a distribution function;
	\item [$P(y_\lambda, y_0, \Delta x)$:] the distribution of $y_\lambda$, with $y_0$ and $\Delta x$ its parameters, given in equation~(\ref{eq:P_lambda});
	\item [$P(y_\theta, y_0, \Delta x)$:] the distribution of $y_\theta$, with $y_0$ and $\Delta x$ its parameters, given in equation~(\ref{eq:P_theta});
	\item [$P(\varphi,x)$:] the distribution of $\varphi$, with $x$ its parameter, given in equation~(\ref{eq:P_phi1});
	\item [$P(\ln\lambda)$:] the stable distribution of $\ln\lambda$, which is a normal distribution with $\overline{\ln\lambda}$ its mean and $\sigma_\lambda$ its standard deviation;
	\item [$P(\theta)$:] the stable distribution of $\theta$, which is a normal distribution with $\bar{\theta}$ its mean and $\sigma_\theta$ its standard deviation.
\end{description}

Below is the pseudo-code:

\begin{algorithmic}
	\State $halo \gets tree\cdot firstHalo$
	\While{$halo \ne null$}
	\If{$halo\cdot child \ne null$}
	\State $child\_halo\gets halo\cdot child$
	\State $\Delta x\gets \ln [child\_halo\cdot mass/halo\cdot mass]$
	\State $y_0\gets [\ln (child\_halo\cdot \lambda)-\overline{\ln\lambda}]/\sigma_\lambda $
	\State $y_\lambda\gets P(y_\lambda, y_0, \Delta x)\cdot sample$
	\State $halo: \lambda\gets\exp(y_\lambda\times \sigma_\lambda+\overline{\ln\lambda})$
	\State $y_0\gets (child\_halo\cdot \theta-\bar{\theta})/\sigma_\theta$
	\State $y_\theta\gets P(y_\theta, y_0, \Delta x)\cdot sample$
	\State $halo: \theta\gets y_\theta\times \sigma_\theta+\bar{\theta}$
	\State $x_0\gets child\_halo\cdot x+\Delta x$
	\State $halo: \varphi\gets P(\varphi,x_0)\cdot sample$
	\State $halo: x\gets x_0$
	\Else
	\State $halo: \lambda\gets \exp[P(\ln\lambda)\cdot sample]$
	\State $halo: \theta\gets P(\theta)\cdot sample$
	\State $halo: \varphi\gets 0$
	\State $halo: x\gets 0$
	\EndIf
	\State $halo \gets halo\cdot next$
	\EndWhile
\end{algorithmic}

\end{document}